\documentclass{IEEEtran4PSCC}
\usepackage[dvipsnames]{xcolor}

\ifCLASSINFOpdf
   \usepackage[pdftex]{graphicx}
\else
   \usepackage[dvips]{graphicx}
\fi
\usepackage[cmex10]{amsmath}

\usepackage{booktabs} 
\usepackage{multirow}  
\usepackage{float}
\usepackage{cite}
\usepackage{tabularx}
\usepackage{multirow}
\usepackage{array}

\makeatletter
\let\old@ps@headings\ps@headings
\let\old@ps@IEEEtitlepagestyle\ps@IEEEtitlepagestyle
\def\psccfooter#1{%
    \def\ps@headings{%
        \old@ps@headings%
        \def\@oddfoot{\strut\hfill#1\hfill\strut}%
        \def\@evenfoot{\strut\hfill#1\hfill\strut}%
    }%
    \def\ps@IEEEtitlepagestyle{%
        \old@ps@IEEEtitlepagestyle%
        \def\@oddfoot{\strut\hfill#1\hfill\strut}%
        \def\@evenfoot{\strut\hfill#1\hfill\strut}%
    }%
    \ps@headings%
}
\makeatother

\begin{document}
%
\title{Modelling and Simulation of Power Systems with Grid-Connected Converters in OpenModelica}

\author{
\IEEEauthorblockN{Lluc Figueras Llerins\\ Vinícius Albernaz Lacerda\\ Eduardo Prieto-Araujo\\ Oriol Gomis-Bellmunt}
\IEEEauthorblockA{Centre d'Innovació Tecnològica en Convertidors \\
Estatics i Accionaments (CITCEA-UPC)\\
Barcelona, Spain}
\and
\IEEEauthorblockN{Adrien Guironnet\\ Quentin Cossart}
\IEEEauthorblockA{R\&D department\\
RTE Réseau de Transport d’Electricité\\
Paris, France}
}


\maketitle

\begin{abstract}
This paper analyses the capabilities of the OpenModelica environment to perform electromagnetic transient (EMT) type simulations of power transmission systems with grid-connected converters. A power transmission system has been modelled and simulated in OpenModelica and Simulink to compare both tools in terms of accuracy, robustness, flexibility and computational performance. Power system transient studies such as faults and switching of capacitor banks have been performed. The results confirmed an excellent overall agreement between both software and demonstrated a remarkable potential for using OpenModelica for EMT-type modelling and simulation of future power electronic dominated grids.
\end{abstract}


\thanksto{\noindent This project has received funding from the European Union’s Horizon 2020 research and innovation programme under grant agreement No 883985 (POSYTYF project). The work of Oriol Gomis-Bellmunt is supported by the ICREA Academia program. Eduardo Prieto-Araujo is a Serra H\'{u}nter Lecturer.}

\section{Introduction}
Future power systems development and operation require open and flexible simulation environments to be able to deal with the new challenges arising, notably with the large increase of power electronic based components such as wind and photo-voltaic generation as well as high voltage direct current links, microgrids and Dynamic Virtual Power Plants (DVPPs) \cite{IEEEhowto:DVPP,IEEEhowto:entsoe0,IEEEhowto:entsoe1}. An easy and straightforward collaboration between the different power system actors (namely transmission system operators, academics or manufacturers) in such a fast evolving environment is needed to ensure the future system's security. Moreover, the use of simulation tools becomes even more necessary to anticipate all these changes. However, though current tools are very powerful and have proved their efficiency over the last decades, they are generally closed software with their own proprietary format, models and solvers, thus making the collaboration more complex.

The advances done in declarative and equation-based modelling languages, particularly in Modelica, offer a promising and mature enough alternative approach. It appears as an excellent candidate to provide an open standard implementation, easy to develop, understand and share among actors \cite{IEEEhowto:Modelica}.  

As mentioned above, current power systems modelling and simulation tools are very powerful but with their own format and structure. Consequently, the collaboration between different actors is not easy and the models are not transparent, making it more complicated to understand the simulation's fundamental components and behaviour and assess the simulation results in detail. Therefore, instead of imposing a modelling software, a common language can be proposed. Equation-based languages such as Modelica can disconnect the dependency between the power system tool and the power system model and provide an open standard implementation~\cite{IEEEhowto:unambiguous}. Hence, Modelica-based models can be exchanged and validated between different Modelica tools. This approach allows the model and the solver to be entirely independent and decoupled, unlike most proprietary power system simulation tools. Accordingly, numerical methods can also be shared across Modelica tools providing remarkable flexibility.

The simulation of electromagnetic transients has become essential in many power system studies due to the increasing penetration of power electronic devices into the grid. Performing EMT-type studies implicates simulating a power system at a very high accuracy level and in a wideband of frequencies (from 0 Hz to 1 MHz or more) by reproducing the actual time-domain waveforms of state variables in the system \cite{IEEEhowto:emt1, IEEEhowto:emt2, IEEEhowto:emt3}. Thus, fast controls of converters can be represented and the interaction between fast-acting power electronic devices can be studied. In power system analysis, Modelica existing studies have mainly focused on the phasor-domain simulations of electromechanical transients and voltage stability analysis. Modelica libraries such as ObjectStab \cite{IEEEhowto:ObjectStab}, iPSL \cite{IEEEhowto:iPSL}, Dyna$\omega$o \cite{IEEEhowto:Dynawo} and PowerGrids \cite{IEEEhowto:PowerGrids} have been developed. However, there is only limited literature on the application of Modelica to electromagnetic transients simulations \cite{IEEEhowto:ModelicaEMT0,IEEEhowto:ModelicaEMT1,IEEEhowto:ModelicaEMT2}. An electrical library of Modelica from the Modelica Standard Library (MSL) is available although it provides simple models with limitations to perform EMT simulations. Modelica language requires an environment to transform it into executable code and run simulations. Multiple Modelica tools are available, either commercial such as Dymola or open-source such as OpenModelica or Dyna$\omega$o .  

OpenModelica offers a fully open-source modelling and simulation environment based on the Modelica language \cite{IEEEhowto:OpenModelica}. It is a powerful tool that can be used to develop and simulate complex systems. OpenModelica is becoming attractive for modelling and simulating power generation and transmission systems due to its advantages compared to existing tools - transparent, easy to use, linked to different top-of-the-art open-source solvers and supporting model exchange through the Functional Mock-Up Interface (FMI) standard \cite{IEEEhowto:FMI}. The FMI standard is used for model exchange and co-simulation between different software. 

Thus, the models implemented and used in OpenModelica are not locked in the development environment, differently from conventional power system modelling and simulation software. Considering this, with FMI and Modelica, the models can be shared and used in other platforms, which offers a wide range of possibilities.

This paper aims to explore the OpenModelica environment capabilities compared to Simulink for EMT-type simulations of power transmission systems with grid-connected converters. The main contributions of this paper are:
\begin{itemize}
    \item [1.] Assess OpenModelica robustness to perform EMT-type simulations.
    \item [2.] Evaluate the accuracy of OpenModelica.
    \item [3.] Analyse OpenModelica computational aspects to simulate electromagnetic transients of power systems with grid-connected converters.
\end{itemize}
This paper is organised as follows. In Section II, the system of study, its implementation and the employed methodology to conduct the analysis are presented. Section III shows the simulation results along with punctual and detailed discussions. A general discussion regarding the numerical performance of both software is provided in Section IV. Finally, in Section V, the conclusions are drawn and future work is outlined.

\section{Methodology}
This section presents the system of study, the implementation of the models and the comparison methodology.
\subsection{Test system}
\begin{figure*}[t]
	\vspace{-3pt}
	\centering
	\includegraphics[width=0.78\textwidth]{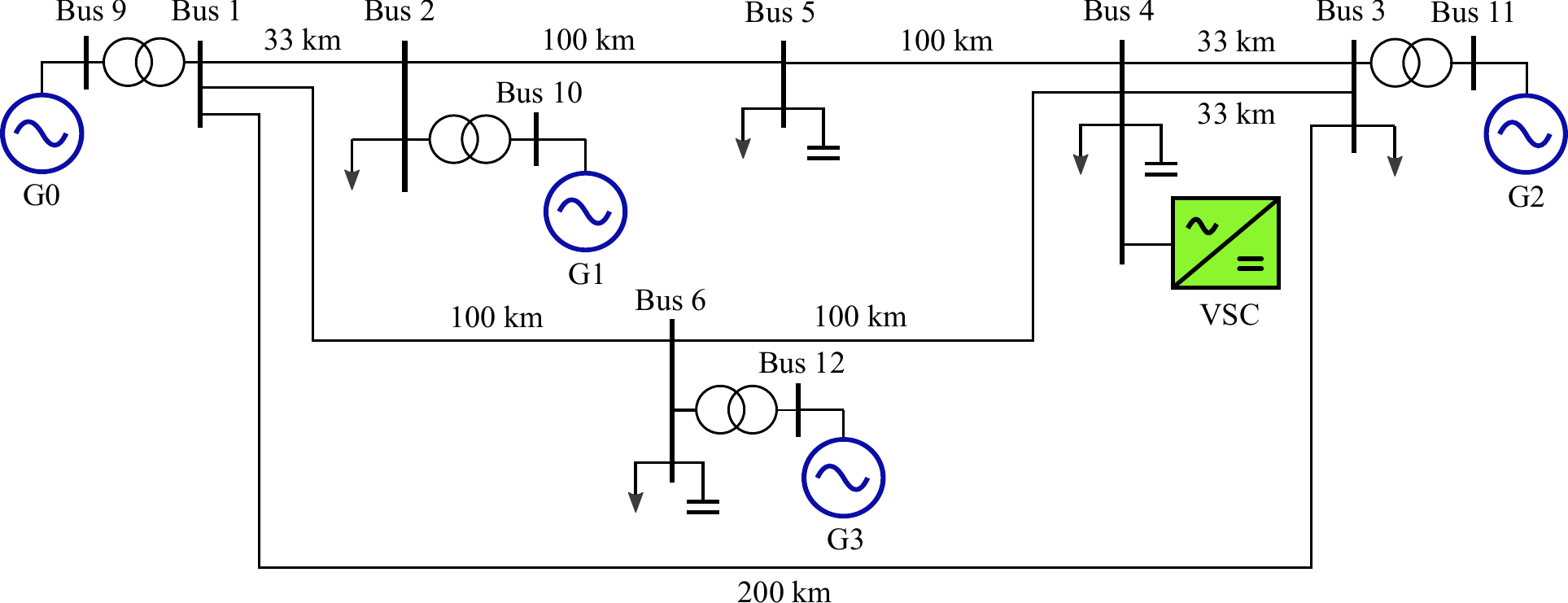}
	\vspace{-3pt}
	\caption{Network single-line diagram. Modified from \cite{IEEEhowto:Cigre}.}
	\vspace{-10pt}
	\label{fig_large_system_diagram}
\end{figure*}

An adaptation of the CIGRE European HV transmission network benchmark system \cite{IEEEhowto:Cigre} with penetration of power electronic converters has been implemented in OpenModelica. The system consists of 12 buses, 4 synchronous generators with transformers, 9 transmission lines, 5 loads and 3 shunt capacitors as shown in Fig.~\ref{fig_large_system_diagram}. A 100 MVA Two-Level Voltage Source Converter (2L-VSC) is connected to bus 4 to conveniently study and analyse the incorporation of large-scale renewable energy sources such as solar or wind energy conversion into the grid. All the models implemented are three-phase EMT-type. Transmission lines are modelled as PI sections and loads as parallel resistive and inductive constant impedances. The implemented Modelica model of this 12-buses system contains 4320 equations and unknowns.

\subsection{Two-level VSC Model}
\begin{figure}[H]
	\vspace{-3pt}
	\centering
	\includegraphics[width=0.48\textwidth]{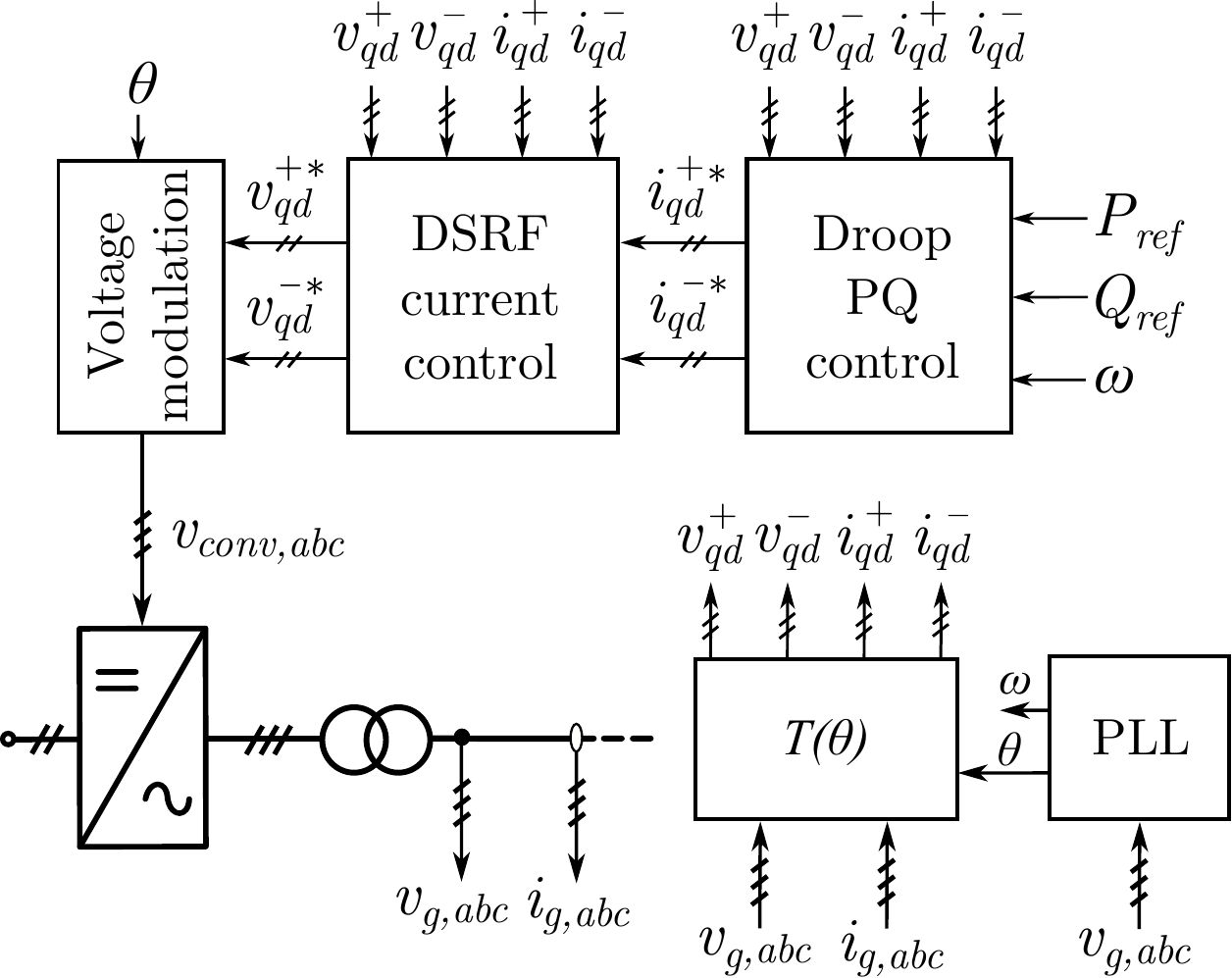}
	\vspace{-3pt}
	\caption{VSC converter and control structure.}
	\label{fig_control_structure}
\end{figure}

The implemented 2L-VSC is an average value model with a classic hierarchical control structure, as shown in Fig~\ref{fig_control_structure}. The current control loop regulates the positive- and negative-sequence current references in $dq$ frame using the Double Synchronous Reference Frame (DSRF) \cite{IEEEhowto:DSRF}. The Delayed Signal Cancellation (DSC) technique \cite{IEEEhowto:DSC} has been chosen to implement the DSRF. This technique is based on a combination of positive- and negative-sequence component vectors and allows achieving accurate information on sequence components with a time delay of one-quarter of a period (5 ms at 50 Hz). The droop PQ control loop regulates the AC power injected into the grid \cite{IEEEhowto:PQcontrol}. The conventional frequency droop without deadband has been implemented, where the additional active power injected by the converter into the grid is proportional to the frequency deviation. Likewise, a conventional voltage droop is implemented, where the extra reactive power injected into the grid is proportional to the voltage deviation. A Low Voltage Ride Through (LVRT) characteristic has been implemented to inject reactive current to support the grid during and after voltage sags and faults \cite{IEEEhowto:LVRT}. A Phase-Locked Loop (PLL) is used to track the angle, frequency and voltage of the AC grid \cite{IEEEhowto:pll}. Proportional-integral controllers set to follow first-order dynamics are used both in the PQ loop and current loop \cite{IEEEhowto:Teodorescu}.

\subsection{Synchronous Generator Model}
This paper implements the classical $dq$ model of a balanced wye-grounded Synchronous Generator (SG). The SG includes the field winding effects, two damper windings on the quadrature axis and an additional damper winding on the direct axis, considering the IEEE model 2.2 with modelled saturation \cite{IEEEhowto:IEEE1110}. The implementation of the SG model in Modelica is based directly on its equations, and they are linked with the main network equations through the $dq0$ transformation. The standard parameters are used as input in the Modelica model, and then the fundamental machine parameters are calculated using the classical definition \cite{IEEEhowto:Kundur}. The generators are connected to the grid by YNd11 (22/220 kV) step-up transformers and their control consists of governor type IEEE TGOV1, exciter type IEEE SEXS and power system stabilizer type IEEE PSS2A \cite{IEEEhowto:ENTSOE}. The synchronous generators parameters are shown in Table~\ref{tab_SG_large_system}.

\begin{table}[h]
    \vspace{-6pt}
	\centering
		\caption{Synchronous generators parameters}
	\begin{tabular}{@{\hspace{1\tabcolsep}}lllll@{\hspace{1\tabcolsep}}}
		\toprule
		& $G0$    & $G1$    & $G2$    & $G3$    \\ \midrule
		$S_n$ (MVA)  & 1000  & 700   & 500   & 500   \\
		$V_n$ (kV)   & 22  & 22.0  & 22.0  & 22.0  \\
		$X_l$ (pu)   & 0.15  & 0.15  & 0.15  & 0.15  \\
		$R_s$ (pu)   & 0.01  & 0.01  & 0.01  & 0.01  \\
		$X_d$ (pu)   & 2  & 2  & 2.2 & 2.2  \\
		$X_d'$ (pu)  & 0.349  & 0.349 & 0.295  & 0.295 \\
		$T_{d}'$ (s)  & 0.935  & 0.935  & 1.002 & 1.002    \\
		$X_d''$ (pu) & 0.225  & 0.225 & 0.193 & 0.193 \\
		$T_{d}''$ (s)  & 0.026  & 0.026  & 0.03  & 0.03  \\
		$X_q$ (pu)   & 1.8  & 1.8  & 2 & 2  \\
		$X_q'$ (pu) & 0.486  & 0.486 & 0.368 & 0.368 \\
		$T_{q}'$ (s) & 0.654  & 0.654 & 0.332 & 0.332 \\
		$X_q''$ (pu) & 0.3  & 0.3 & 0.2 & 0.2 \\
		$T_{q}''$ (s) & 0.046  & 0.046 & 0.022 & 0.022 \\
		$H$ (s)  & 6   & 5   & 3   & 5 \\ \bottomrule
	\end{tabular}
\label{tab_SG_large_system}
\end{table}

\subsection{System Initialization}
To initialize a Modelica model, initial guess values for all discrete and continuous variables need to be provided to make initialization stable. Since there is index reduction going on, all of the states in the system must be initialized adequately \cite{IEEEhowto:init0}. These values should be entered manually or computed through an external routine. 
Dynamic power grid models require the results of a static load flow computation to set up the start values for initialization. As OpenModelica lacks a power flow computation tool, voltage for lines and loads and initial values for the generator models, governors, voltage regulators and power system stabilizers are obtained from the steady-state solution from the power flow computed in Simulink. Therefore, the process of model comparison becomes reliable since both Modelica and Simulink have the same initial conditions. Notwithstanding, previous studies have introduced Modelica tools to link time-domain simulations with static computations like power flows since they are crucial to perform power system analysis \cite{IEEEhowto:init1,IEEEhowto:init2,IEEEhowto:init3}. Modelica libraries such as PowerGrids \cite{IEEEhowto:PowerGrids} include models built for this purpose.
In the studied test case, Modelica synchronous generators and PI section lines models require explicit initialization. Thus, Modelica $initialization~models$ are used instead of $initial~equation$ inside the models (the common initialization strategy in Modelica). This considerably reduces the size of the initial system to be solved and increases simulation speed (notably for large systems) \cite{IEEEhowto:init4}. 

With the particular initialization scheme adopted in this Modelica system, flat initialization is not achieved. For that reason, to simulate the events, both systems need to achieve steady-state. After the VSC starts injecting active and reactive power at $t$~=~1~s and $t$~=~1.2~s, respectively, the tests results are anlysed at $t$~=~50~s to ensure the system is in steady-state condition. Therefore, all the time instants referred to in the simulation results section are after $t$~=~50~s. 

\subsection{Comparison Methodology}
The models have been tested and compared against Simulink as the reference tool. Four types of events are simulated to study the system behaviour: $i$) a load connection; $ii$) a three-phase symmetrical fault; $iii$) a single line-to-ground fault; $iv$) a capacitor bank switching. These events represent essential studies in EMT simulations of systems with power electronic penetration and they allow analysing the interaction between the power converter and the grid. The results obtained with OpenModelica are compared against the same model in Simulink. Comparisons between OpenModelica and Simulink accuracy and performance have been performed. 

Numerical tests are performed using the variable-step DASSL ODE solver with a maximum step size of 25~$\mu$s, tolerance of 1e-4 and the maximum integration order of 5 in OpenModelica. DASSL integration method is based on the Backward Differentiation Formula (BDF) \cite{IEEEhowto:dassl}. In Simulink, a Backward Euler integration method with a step size of 25~$\mu$s is employed. All simulations run on a desktop with AMD Ryzen Threadripper 2950X 16-Core 3.50 GHz Processor and 32 GB of RAM under Windows 10 64-bit.

\section{Simulation results}
This section presents the simulation results of the modified CIGRE European HV transmission network benchmark system \cite{IEEEhowto:Cigre} to compare Simulink and Modelica models. The VSC behaviour and its interaction with the grid during the events are analysed.

\subsection{Load connection}
\begin{figure}[h]
	\centering
	\includegraphics[width=0.47\textwidth]{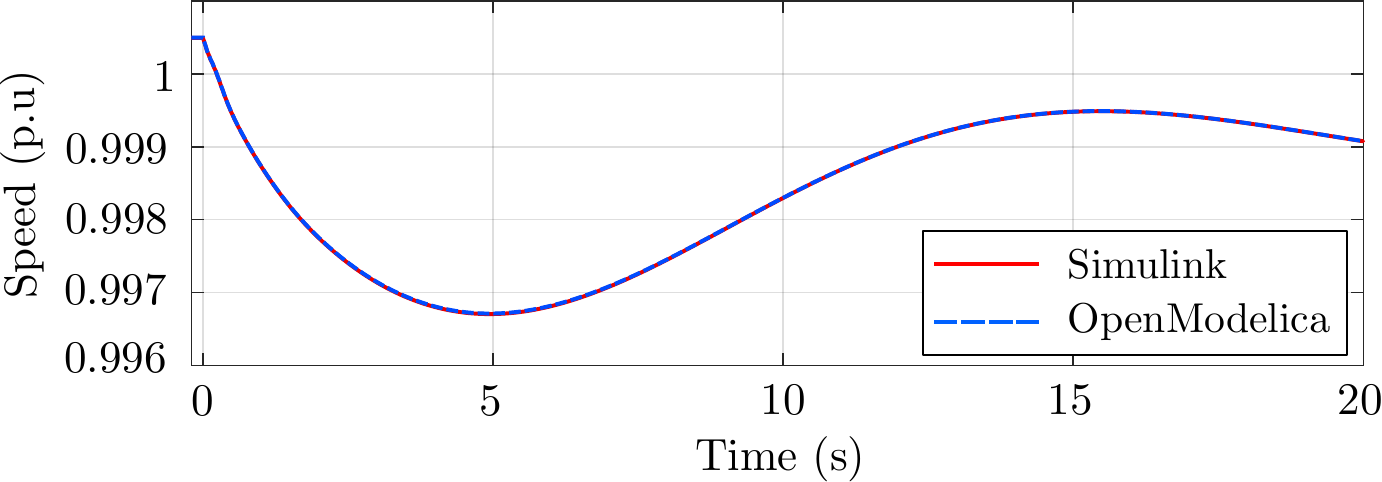}
	\vspace{-5pt}
	\caption{Speed response of G1 during a load connection.}
	\vspace{-5pt}
	\label{fig_fvdev_omega}
\end{figure}

\begin{figure}[h]
	\vspace{-5pt}
	\centering
	\includegraphics[width=0.47\textwidth]{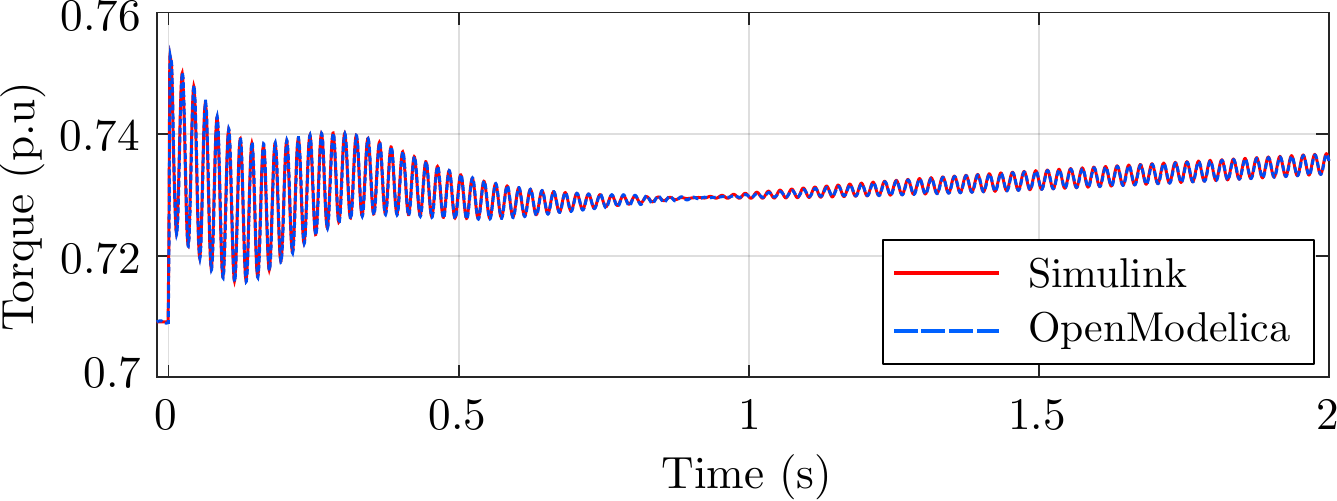}
	\vspace{-5pt}
	\caption{Electrical torque response of G1 during a load connection.}
	\vspace{-5pt}
	\label{fig_fvdev_te}
\end{figure}

In this test, an instantaneous increase in the load at bus 5 of 100~MW and 30~Mvar takes place at  $t$~=~100~ms.
A comparison between the simulation results of both models is presented in Fig.~\ref{fig_fvdev_omega} and Fig.~\ref{fig_fvdev_te}. Focusing on G1 behaviour, the speed and the electrical torque responses show a perfect match between the two implementations. Therefore, the results guarantee consistent simulation of electromechanical transients.

\subsection{Symmetrical fault}
\begin{figure}[ht]
	\vspace{-5pt}
	\centering
	\includegraphics[width=0.47\textwidth]{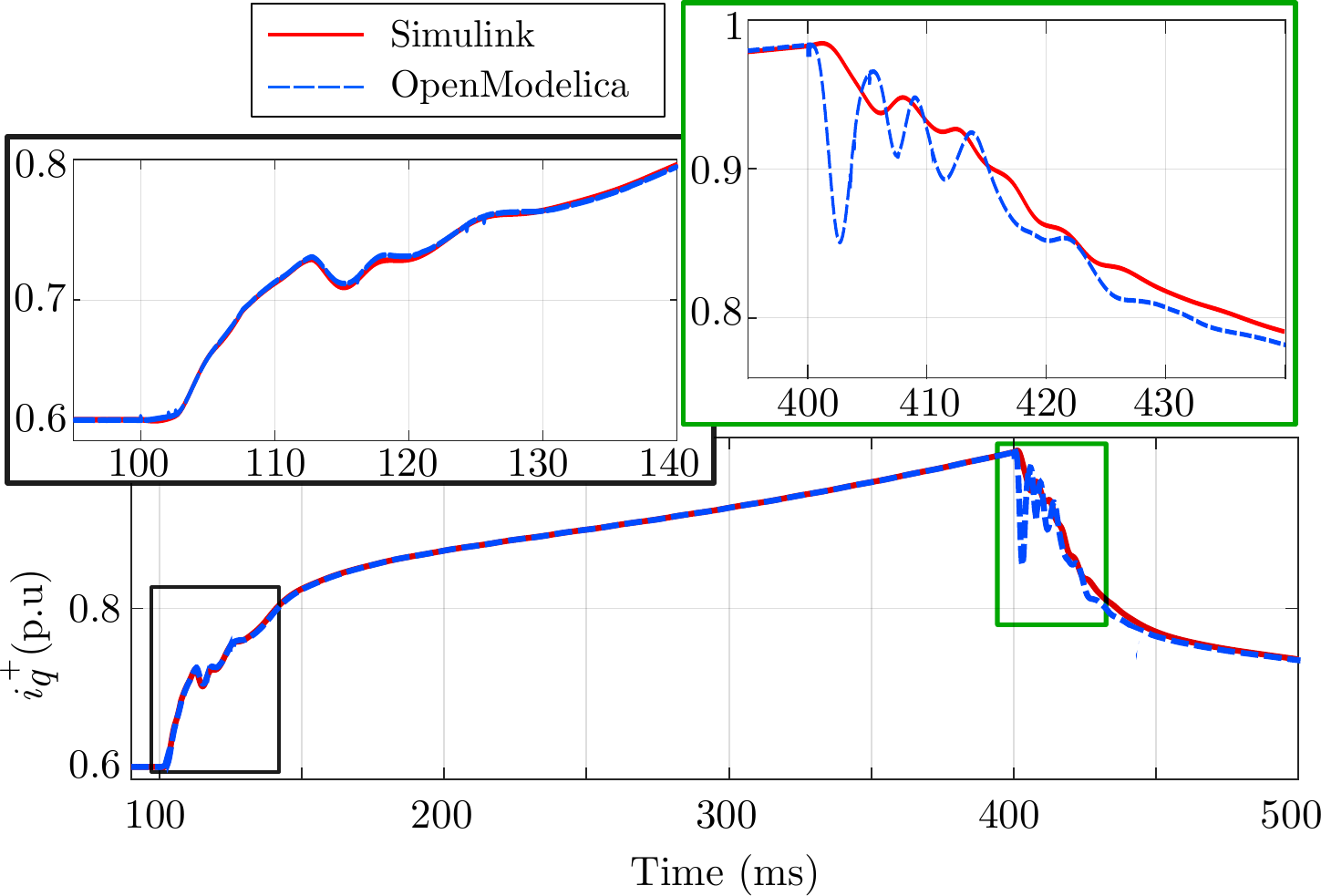}
	\vspace{-5pt}
	\caption{VSC $i_{q}^{+}$ current response during a three-phase fault on bus 6.}
	\vspace{-5pt}
	\label{fig_sf_iqpos}
\end{figure}

\begin{figure}[ht]
	\vspace{-5pt}
	\centering
	\includegraphics[width=0.47\textwidth]{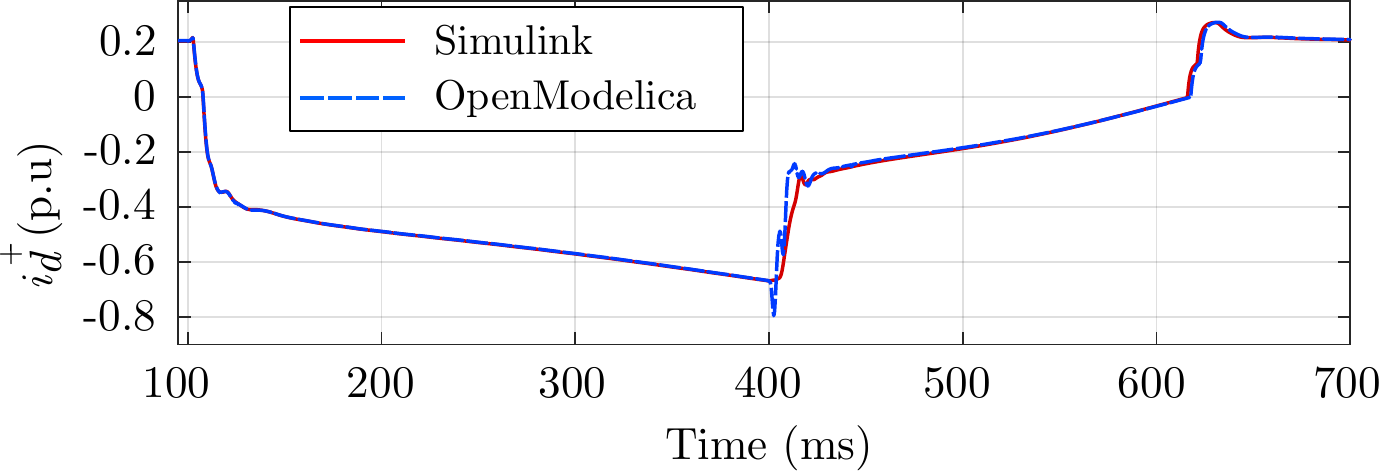}
	\vspace{-5pt}
	\caption{VSC $i_{d}^{+}$ current response during a three-phase fault on bus 6.}
	\vspace{-5pt}
	\label{fig_sf_idpos}
\end{figure}

\begin{figure}[H]
	\vspace{-5pt}
	\centering
	\includegraphics[width=0.47\textwidth]{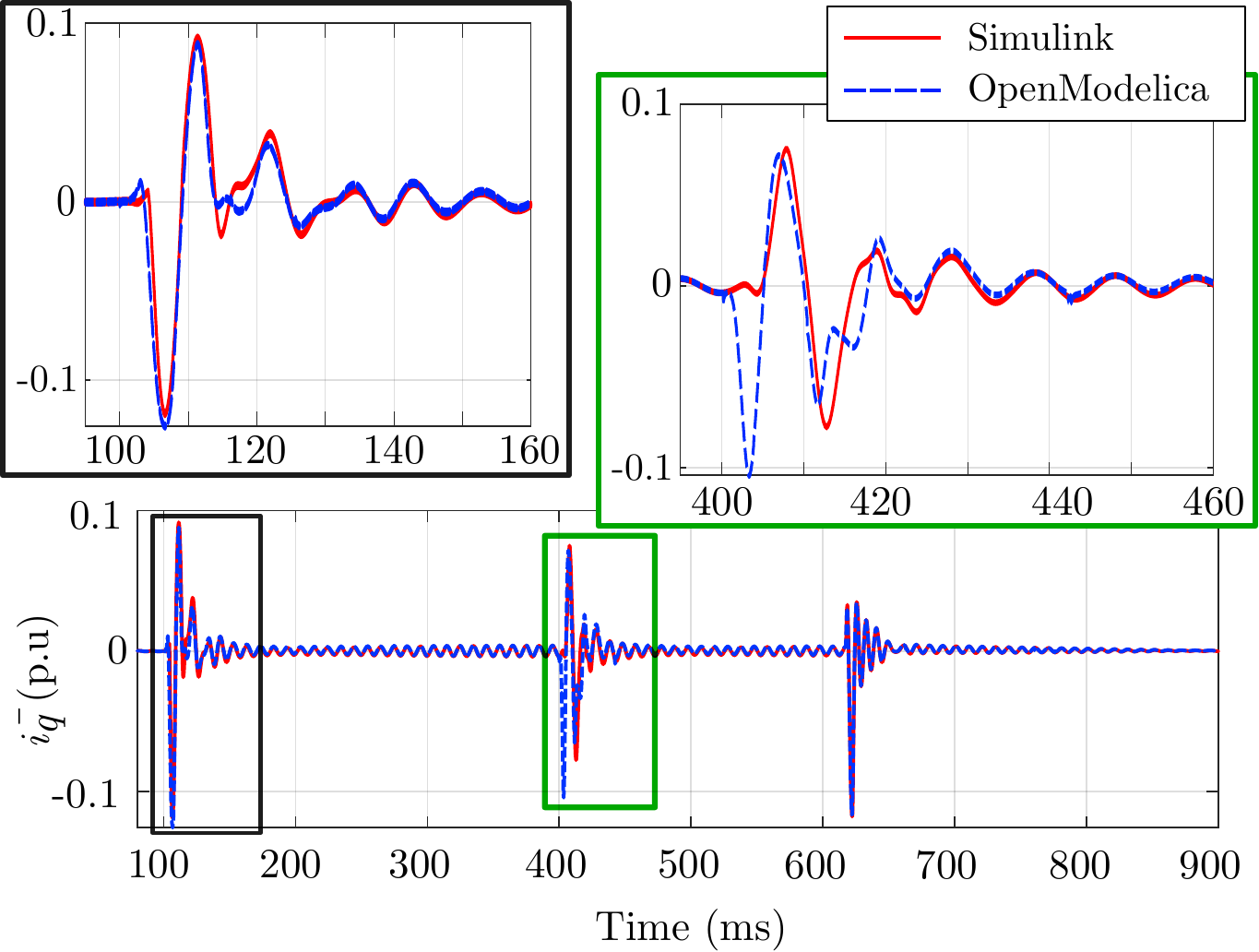}
	\vspace{-5pt}
	\caption{VSC $i_{q}^{-}$ current response during a three-phase fault on bus 6.}
	\vspace{-5pt}
	\label{fig_sf_iqneg}
\end{figure}

\begin{figure}[ht]
	\vspace{-5pt}
	\centering
	\includegraphics[width=0.47\textwidth]{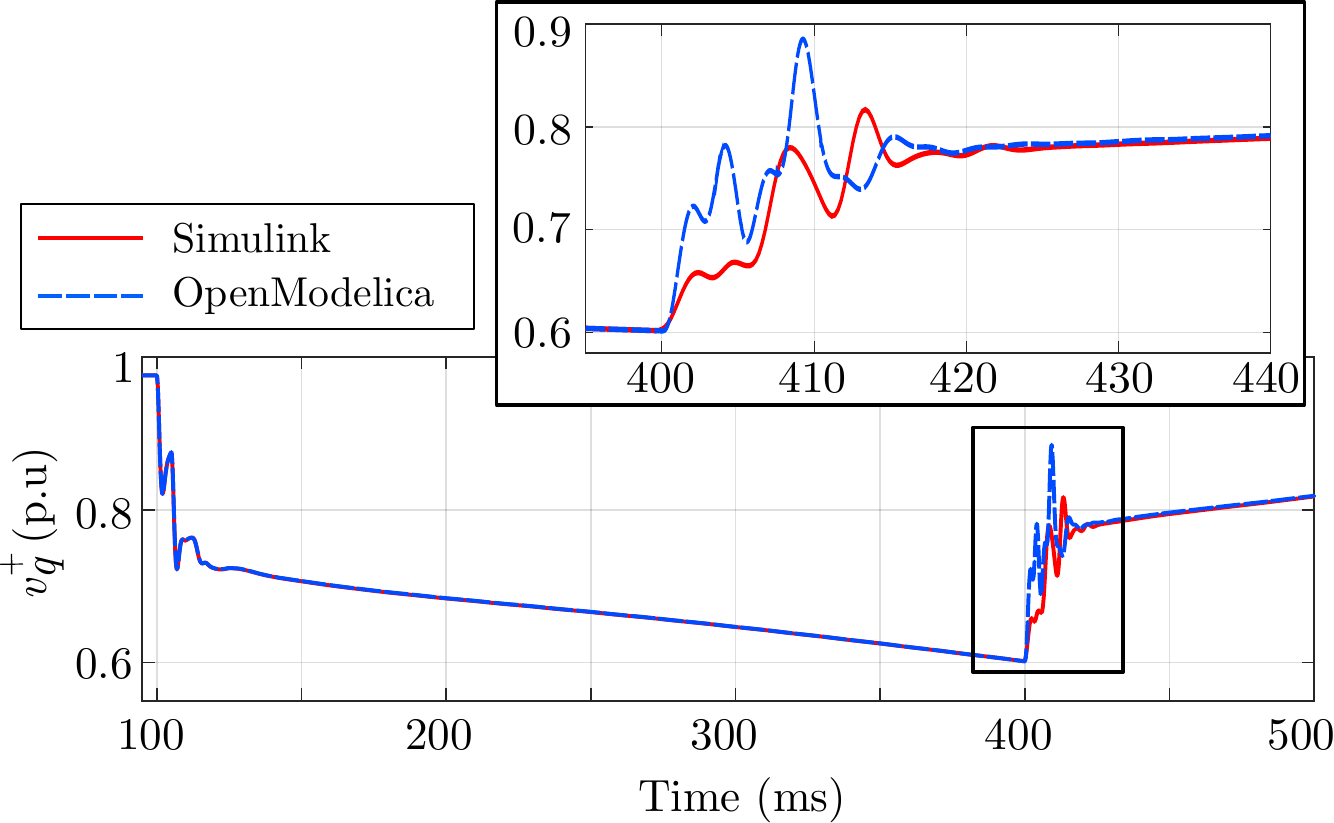}
	\vspace{-5pt}
	\caption{VSC  $v_{q}^{+}$ voltage response during a three-phase fault on bus 6.}
	\vspace{-5pt}
	\label{fig_sf_vqpos}
\end{figure}

\begin{figure}[ht]
	\vspace{-5pt}
	\centering
	\includegraphics[width=0.47\textwidth]{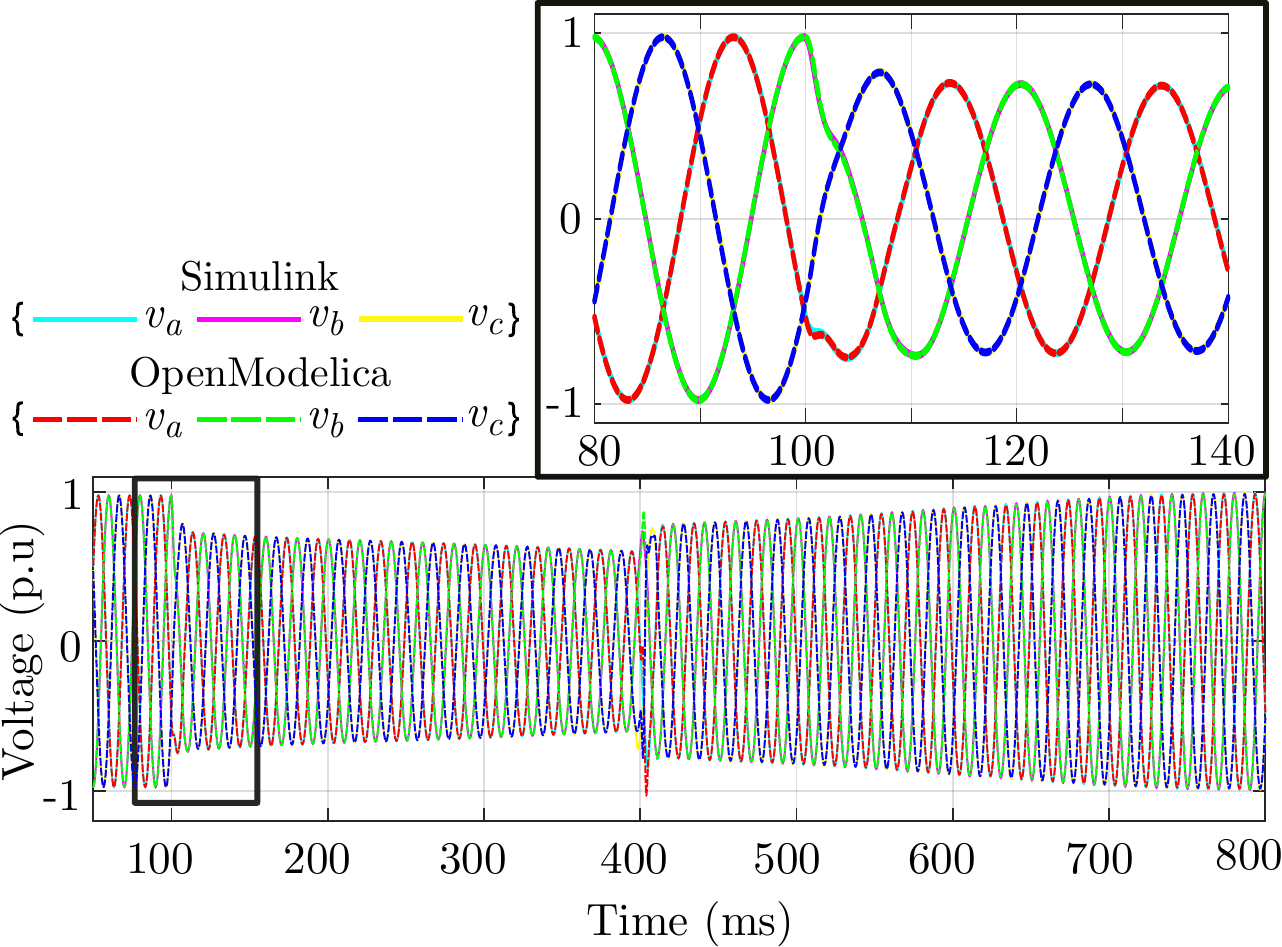}
	\vspace{-5pt}
	\caption{Three-phase voltages in the VSC terminal during a three-phase fault on bus 6.}
	\vspace{-5pt}
	\label{fig_sf_vabc}
\end{figure}

In this test, a 5 $\Omega$ three-phase fault lasting 300~ms is applied on bus 6 at $t$~=~100~ms. The VSC negative-sequence control and the LVRT are enabled during this test. Fig.~\ref{fig_sf_iqpos}, Fig.~\ref{fig_sf_idpos} and Fig.~\ref{fig_sf_iqneg} show $i_{q}^{+}$,  $i_{d}^{+}$ and $i_{q}^{-}$ converter current responses, respectively.  Fig.~\ref{fig_sf_vqpos} depicts the converter positive-sequence voltage $v_{q}^{+}$ while Fig.~\ref{fig_sf_vabc} shows and the converter terminal three-phase voltages waveforms. 

Before the fault, the VSC delivers 60~MW active power and 20~Mvar reactive power. When the fault occurs, the positive-sequence currents $i_{q}^{+}$ and $i_{d}^{+}$ initial transient responses are nearly identical in both simulators. When the fault is cleared, a mismatch can be observed during the following 40~ms, where the Modelica model presents a larger oscillatory component than Simulink. 
The negative-sequence current $i_{q}^{-}$ is kept close to 0 during and after the fault. When the fault takes place and when is removed, one can observe minor deviations between both responses.
Analysing the three-phase voltages at the converter terminal and the positive-sequence voltage $v_{q}^{+}$ responses can be observed that the VSC is able to sustain the terminal voltage at a relatively high level during the fault and exhibits a stable response after the fault is cleared. The LVRT is enabled due to a voltage drop lower than 0.9~pu at $t$~=~107~ms to supply reactive power to the grid. After the fault, when the voltage is restored at $t$~=~516~ms, the LVRT is automatically disabled. When the fault is cleared at $t$~=~400~ms, the oscillatory component in the $v_{q}^{+}$ is more significant again in OpenModelica. The voltage at the converter terminal recovers the pre-fault values at $t$~=~800~ms.
These discrepancies in the oscillatory component of the responses when the fault is cleared are due to the use of different numerical solvers in both simulators, which significantly influences the phase sequence separation. The step size used by DASSL is reduced to 0.12~$\mu$s at the time instant when the fault is cleared. Thus, high-frequency transient oscillations are not captured by Simulink as  $\Delta t$~=~25~$\mu$s.

\begin{figure}[H]
	\vspace{-5pt}
	\centering
	\includegraphics[width=0.47\textwidth]{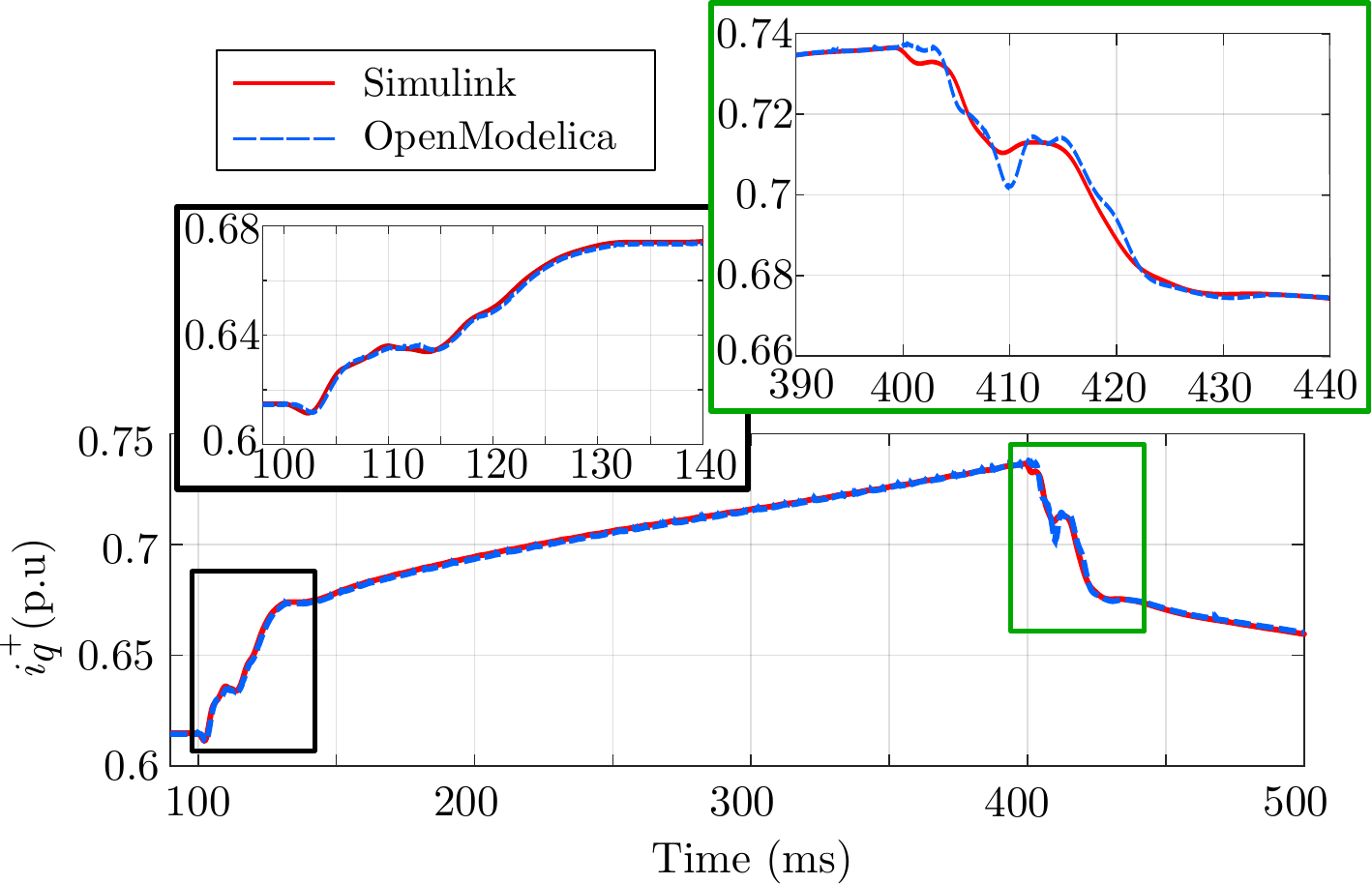}
	\vspace{-5pt}
	\caption{VSC $i_{q}^{+}$ current response during a single line-to-ground fault on~bus~6.}
	\label{fig_af_iqpos}
\end{figure}

\subsection{Asymmetrical fault}

\begin{figure}[H]
	\centering
	\includegraphics[width=0.47\textwidth]{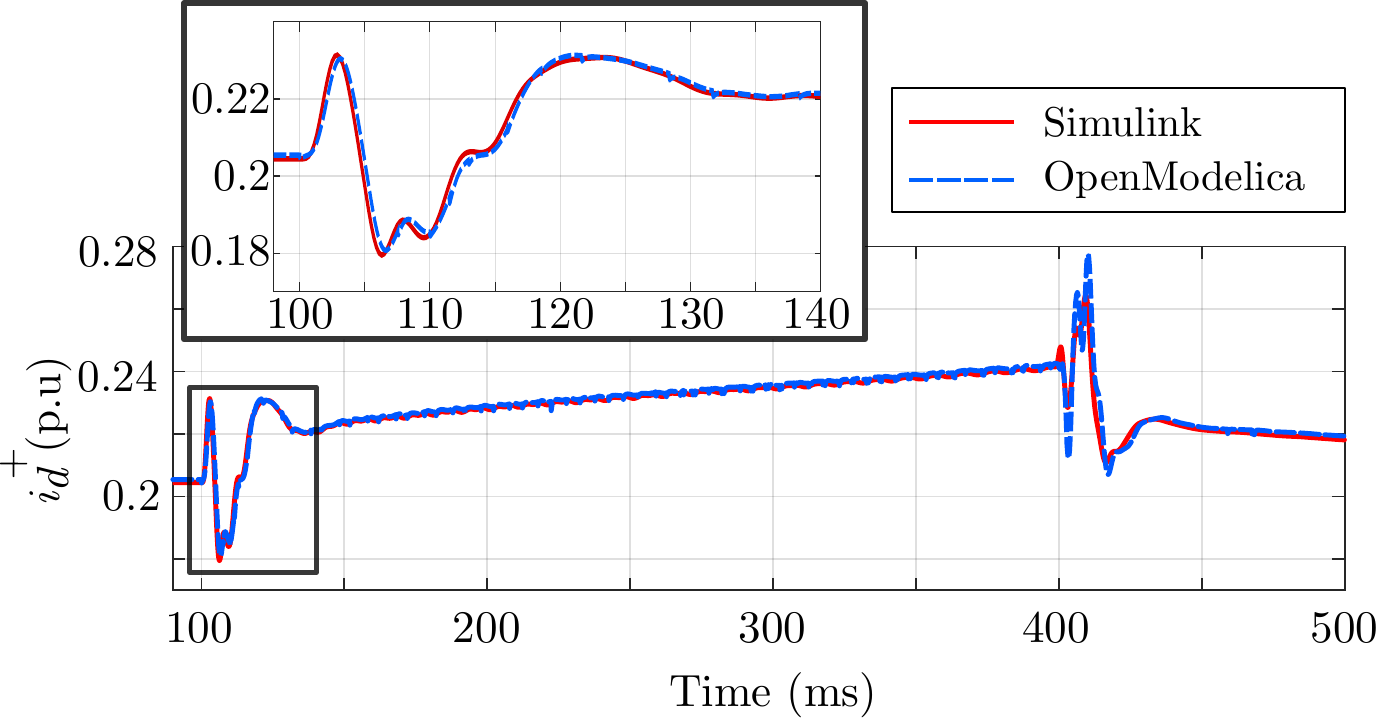}
	\vspace{-6pt}
	\caption{VSC $i_{d}^{+}$ current response during a single line-to-ground fault on~bus~6.}
	\vspace{-5pt}
	\label{fig_af_idpos}
\end{figure}

\begin{figure}[H]
	\vspace{-6pt}
	\centering
	\includegraphics[width=0.47\textwidth]{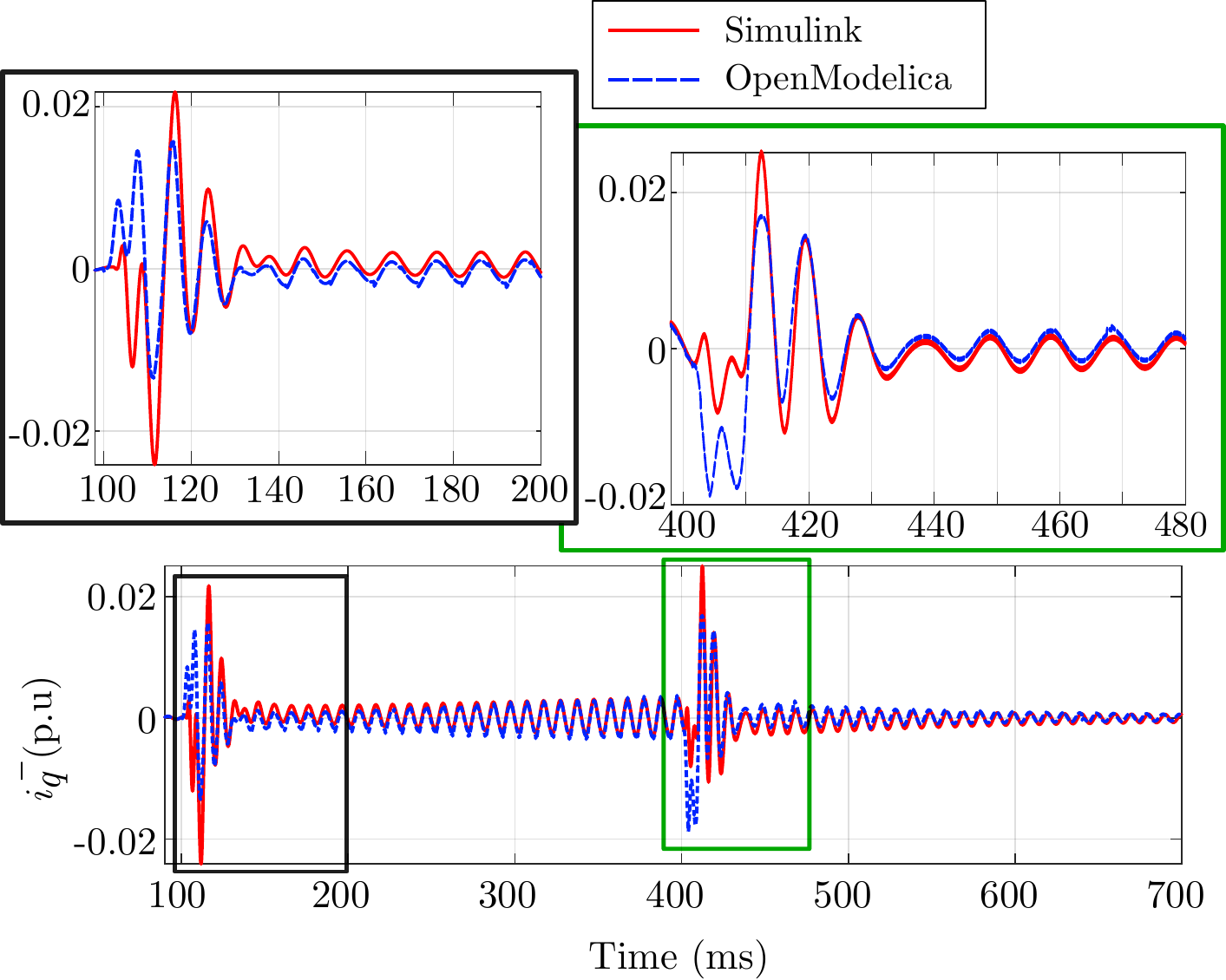}
	\vspace{-6pt}
	\caption{VSC $i_{q}^{-}$ current response during a single line-to-ground fault on~bus~6.}
	\vspace{-5pt}
	\label{fig_af_iqneg}
\end{figure}

\begin{figure}[H]
	\vspace{-6pt}
	\centering
	\includegraphics[width=0.47\textwidth]{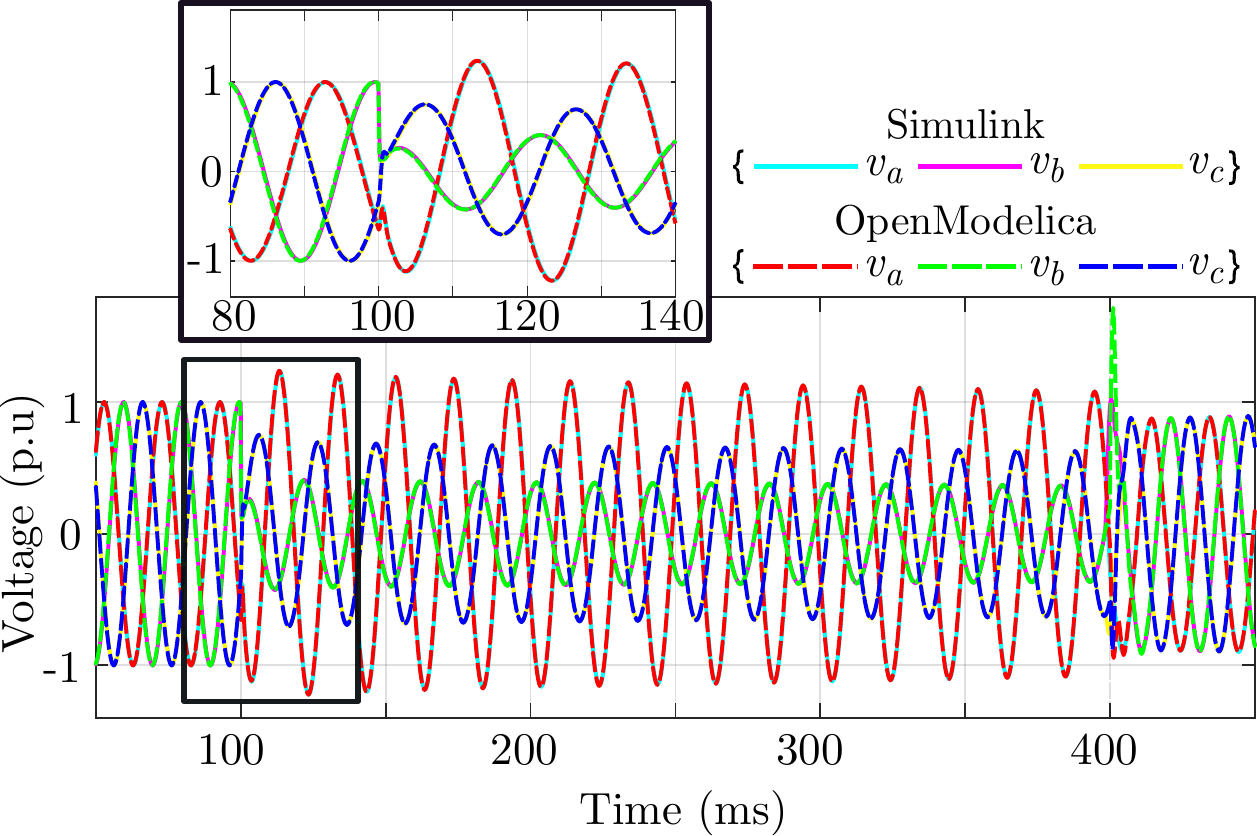}
	\vspace{-6pt}
	\caption{Three-phase voltages in the G3 terminal during a single line-to-ground fault on bus 6.}
	\vspace{-6pt}
	\label{fig_af_vabc}
\end{figure}

In this test, a 5~$\Omega$ single line-to-ground fault at phase $b$ is applied on bus 6 at $t$~=~100~ms, lasting 300~ms. The converter negative-sequence control is enabled during this test. Fig.~\ref{fig_af_iqpos}, Fig.~\ref{fig_af_idpos} and Fig.~\ref{fig_af_iqneg} show $i_{q}^{+}$,  $i_{d}^{+}$ and $i_{q}^{-}$ converter current responses, respectively. Fig.~\ref{fig_af_vabc} depicts the three-phase voltages waveforms at the G3 terminal.
Prior to the fault, the VSC is injecting 60~MW active power and 20~Mvar reactive power into the grid. When the fault occurs, the converter positive-sequence current responses and the three-phase voltages waveforms at the G3 terminal show an excellent agreement during the 40~ms initial transient. When the short-circuit is cleared at $t$~=~400~ms, minor deviations are observed both in the converter positive- and negative-sequence responses. As mentioned in the symmetrical fault scenario, these discrepancies in the oscillations during the fault recovery are due to the different integration methods used in both tools.


\subsection{Capacitor bank switching}
In this test, a capacitor bank switching event is simulated per phase. A 100~Mvar capacitor bank is connected to bus 5. The bank switches close phases $a$, $b$ and $c$ at $t$~=~101~ms, $t$~=~109~ms and $t$~=~117~ms, respectively. In this particular case, the step size used in Simulink is reduced to 5~$\mu$s to avoid missing any dynamics as a fixed-step size solver is employed. It is noted that a close match is achieved between both simulators, even considering the transient due to delayed switching between phases. These results confirm the capability of OpenModelica to simulate unbalanced loads in three-phase systems accurately.  

\begin{figure}[H]
    \vspace{-3pt}
	\centering
	\includegraphics[width=0.47\textwidth]{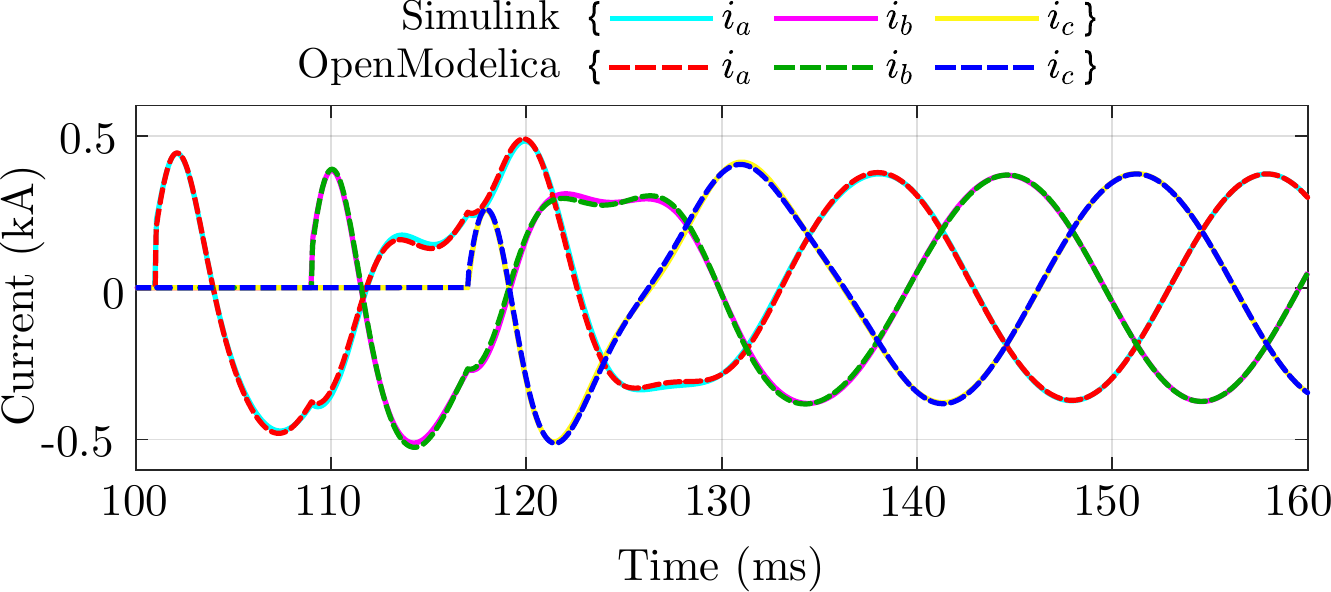}
    \vspace{-4pt}
	\caption{Three-phase capacitor bank inrush currents during a capacitor bank switching.}
	\label{fig_cs_iabc}
\end{figure}

\section{Numerical results and discussion}
This section presents the computational and efficiency comparison between OpenModelica (v1.18.0) environment and Simulink (R2021a) for EMT simulation of power transmission systems with grid-connected converters. 

In OpenModelica, DASSL solver is used due to its accuracy and stability \cite{IEEEhowto:dassl} while in Simulink the Backward Euler integration method is employed. The numerical statistics of OpenModelica simulation of the symmetrical fault test case simulation are presented in Table~\ref{tab_sim_performance}. The simulation is run for 70~s. 

As shown in Table~\ref{tab_sim_performance}, using DASSL with a maximum step size of 25 $\mu$s and a tolerance of 1e-4, the total simulation runtime is 25449~s for a 70~s simulation, while in Simulink is 2133~s using the Backward Euler solver. The Jacobian evaluations have a significant share in the total simulation time of the Modelica system due to the costly Jacobian evaluation as an automatic differentiation algorithm does it and not symbolically derived. Additionally, there is a Jacobian refactorization in DASSL each time the time step slightly varies, which is very frequent with oscillating variables, as variable time step increase is by nature limited with oscillations. 
In order to address this issue, efforts have been made by developing the Sinusoidal Predictor Method (SPM) \cite{IEEEhowto:sinpredmethod}. In EMT simulations, the mathematical problem to solve is a set of differential-algebraic equations. The solution of this set of equations is assumed to contain oscillating and non-oscillating components. Therefore, the idea of the SPM is to exploit these oscillating components' behaviour in steady-state to accelerate the computation. The SPM is implemented into IDA \cite{IEEEhowto:ida}. At the moment, using the SPM leads to a significant reduction of iterations for performing a simulation without loss of accuracy. However, the current implementation of the method is not entirely optimized even though it preserves very high modelling flexibility. Likewise, tolerance has a considerable impact on the computation time. In the tested system, tolerance can be lowered to 1e-3 without a significant loss of accuracy. Minor differences would be noticed only in the fastest dynamics from the converter DSRF current control.

In Modelica-based tools, many algorithms rely on the efficient provision of Jacobian matrices. Accordingly, speeding up the evaluation of Jacobians would lead to a considerable reduction of the total simulation runtime for a broad range of Modelica models. In OpenModelica, efforts are being made in this direction to provide an~analytically determined Jacobian matrix to accelerate computation~\cite{IEEEhowto:jacobian}. 

On the other hand, if the VSC is disconnected from the grid, the simulation runtime decreases by a factor of 25 in OpenModelica as the steps taken by the solver, the calls of ODE functions and the Jacobian evaluations reduce dramatically (see Table~\ref{tab_sim_performance}). This is due to the continuous and high frequency oscillations in the VSC control structure. As aforementioned, there is a refactorization of the Jacobian each time the time step slightly fluctuates, common in oscillating variables. In order to address this issue, alternatives in the VSC implementation and way of modelling could be examined. As shown in Table~\ref{tab_sim_performance}, in the test case with the VSC disconnected, OpenModelica outperforms Simulink in terms of total simulation runtime with the ratio of 2.15:1.

Nonetheless, it is important to mention that the integration methods used in Modelica and Simulink are essentially different, and it is not fair to compare variable-step and fixed-step types of solvers.

\begin{table}[h]
\vspace{-5pt}
\centering
\caption{Simulation performance in OpenModelica}
\label{tab_sim_performance}
\begin{tabular}{|p{2.8cm}|>{\centering\arraybackslash}m{2.2cm}|>{\centering\arraybackslash}m{2.2cm}|}
\hline
\textbf{Characteristics} & \textbf{System with connected VSC} & \textbf{System with disconnected VSC} \\ \hline
Solver                   &         DASSL              &        DASSL           \\ \hline
Tolerance            &     1e-4              &   1e-4            \\ \hline
$\Delta t_{max}$         &           25 $\mu$s           &     25 $\mu$s            \\ \hline
$\Delta t_{min}$         &           0.12 $\mu$s           &    0.13 $\mu$s            \\ \hline
Steps taken            &     39,056,194                 &   778,739               \\ \hline
ODE function calls      &    73,551,404 &     1,537,703              \\ \hline
J-evaluations            &     14,188,205                  &   183,749           \\ \hline
J-evaluation time (s)            &      11,428                &           131       \\ \hline
Total simulation time (s)            &      25,449           &           992       \\ \hline
\end{tabular}
\vspace{-5pt}
\end{table}

\section{Conclusions}
In this paper, the performance of OpenModelica in simulating EMT models of electrical power systems with grid-connected converters has been assessed. OpenModelica has been compared with Simulink in terms of robustness, accuracy, flexibility and computational aspects. 
The results confirmed a notable overall agreement between OpenModelica and Simulink, not only for electromechanical transients but also for electromagnetic transients.

OpenModelica environment has demonstrated excellent potential for EMT-type modelling and simulation of future power electronic dominated grids. Modelica implementation of such models is easy and straightforward due to the native properties of the language (declarative and equation-based). In addition, the decoupling of models from numerical solvers offers outstanding flexibility compared to Simulink. 
In general terms, for the studied system with penetration of power electronic converters, OpenModelica computation performance is not satisfactory compared to Simulink at the moment. However, having a symbolic Jacobian in Modelica-based tools would be a game-changer for its operational use.

Future works include testing larger transmission power systems in the OpenModelica environment, implementing initialization routines to avoid explicitly entered power flow data, and analysing systems with a higher share of Renewable Energy Sources (RES).





%

\end{document}